\documentclass[conference, 10pt]{IEEEtran}

\def\BibTeX{{\rm B\kern-.05em{\sc i\kern-.025em b}\kern-.08em
    T\kern-.1667em\lower.7ex\hbox{E}\kern-.125emX}}

\usepackage[ruled,lined]{algorithm2e}
\usepackage{graphicx}
\usepackage{xspace}
\usepackage{subfig}
\usepackage{siunitx}
\usepackage{soul}
\usepackage{booktabs}
\usepackage{multirow}
\usepackage{multicol}
\usepackage{wrapfig}
\usepackage[dvipsnames]{xcolor}

\ifx\figurename\undefined \def\figurename{Figure}\fi
\renewcommand{\figurename}{Fig.}
\renewcommand{\paragraph}[1]{\textbf{#1} }

\newcommand{\Sect}[1]{Sec.~\ref{#1}}
\newcommand{\Fig}[1]{Fig.~\ref{#1}}
\newcommand{\Tbl}[1]{Tbl.~\ref{#1}}

\newcommand{\Alg}[1]{Algo.~\ref{#1}}

\newcommand{\mode}[1]{\underline{\textsc{#1}}\xspace}

\newcommand{\proj}{\textsc{SLTarch}\xspace}
\newcommand{\arch}{\textsc{LTcore}\xspace}
\newcommand{\spcore}{\textsc{SPcore}\xspace}
\newcommand{\tree}{\textsc{SLTree}\xspace}

\newcommand{\RNum}[1]{\uppercase\expandafter{\romannumeral #1\relax}}

\graphicspath{{figs/}}

\begin{document}

\title{\proj: Towards Scalable Point-Based Neural Rendering by Taming Workload Imbalance and Memory Irregularity}


\author{
    Xingyang Li\textsuperscript{1,4,*}, 
    Jie Jiang\textsuperscript{1,4,*}, 
    Yu Feng\textsuperscript{1,2,\dag}, 
    Yiming Gan\textsuperscript{3}, 
    Jieru Zhao\textsuperscript{1}, 
    Zihan Liu\textsuperscript{1}, 
    Jingwen Leng\textsuperscript{1,2}, 
    Minyi Guo\textsuperscript{1,2} \\

    \textsuperscript{1}Shanghai Jiao Tong University, 
    \textsuperscript{2}Shanghai Qi Zhi Institute, 
    \textsuperscript{3}Chinese Academy of Sciences, 
    \textsuperscript{4}Zhiyuan College \\
    \{brucelee\_sjtu, jiang\_jie, y-feng\}@sjtu.edu.cn 
    \textsuperscript{*}Equal contribution. \textsuperscript{\dag}Corresponding author.
}



\maketitle 
\pagestyle{empty} 

\begin{abstract}

Rendering is critical in fields like 3D modeling, AR/VR, and autonomous driving, where high-quality, real-time output is essential. 
Point-based neural rendering (PBNR) offers a photorealistic and efficient alternative to conventional methods, yet it is still challenging to achieve real-time rendering on mobile platforms. 
We pinpoint two major bottlenecks in PBNR pipelines: \textit{LoD search} and \textit{splatting}.
LoD search suffers from workload imbalance and irregular memory access, making it inefficient on off-the-shelf GPUs. 
Meanwhile, splatting introduces severe warp divergence across GPU threads due to its inherent sparsity.

To tackle these challenges, we propose \proj, an algorithm-architecture co-designed framework.
At its core, \proj introduces \tree, a dedicated subtree-based data structure, and \arch, a specialized hardware architecture tailored for efficient LoD search.
Additionally, we co-design a divergence-free splatting algorithm with our simple yet principled hardware augmentation, \spcore, to existing PBNR accelerators.
Compared to a mobile GPU, \proj achieves 3.9$\times$ speedup and 98\% energy savings with negligible architecture overhead.
Compared to existing accelerator designs, \proj achieves 1.8$\times$ speedup with 54\% energy savings.

\end{abstract}


\section{Introduction}

Rendering is fundamental across various domains, including 3D modeling~\cite{xiangli2022bungeenerf, barron2021mip, lindell2022bacon}, augmented and virtual reality (AR/VR)~\cite{chen2023mobilenerf, hu2022efficientnerf, hedman2021baking}, autonomous driving~\cite{zhou2024drivinggaussian, matsuki2024gaussian, yan2024gs}, and many more~\cite{gao2022nerf, chen2024survey}.
High-quality, real-time rendering is essential in these fields to deliver immersive experiences. 

Among the existing rendering techniques, point-based neural rendering (PBNR)~\cite{kerbl20233d, kerbl2024hierarchical, fang2024mini, lee2023compact} has emerged as the hottest topic in the recent two years~\cite{wu2024recent, chen2024survey}.
Unlike conventional rasterization techniques, PBNR leverages Gaussian point primitives, also called Gaussians, with learnable parameters to achieve photo-realistic rendering, while it can provide performance advantages over other recent advanced techniques, such as ray tracing~\cite{pharr2023physically, deng2017toward, pantaleoni2010hlbvh} and neural radiance field~\cite{hu2022efficientnerf, hedman2021baking, mildenhall2021nerf}.

\paragraph{Motivation.} Despite PBNR having its performance advantages over other techniques, PBNR itself is still challenging to completely replace conventional rasterization pipelines on mobile platforms, due to its scalability and runtime performance~\cite{lin2024rtgs, lee2024gscore}. 
On large-scale datasets~\cite{kerbl2024hierarchical, ren2024octree}, PBNR algorithms~\cite{kerbl2024hierarchical, kerbl20233d} barely achieve 15 frame-per-second (FPS) on mobile Ampere GPU on Nvidia Orin SoC~\cite{orinsoc}, far from real-time VR rendering requirements, 60 FPS~\cite {questprospec, visionprospec}.
To address this, we propose, \proj, an algorithm-architecture co-designed system dedicated to real-time rendering in VR.

While numerous accelerators have been proposed recently for PBNR~\cite{lee2024gscore, feng2024potamoi, lee2025vr, li2025uni, ye2025gaussian, he2025gsarch}, they exclusively focus on accelerating one specific stage of PBNR, \textit{splatting}, while often overlooking another crucial stage, \textit{level-of-detail (LoD) search}, in PBNR.
Our key observation is that, as the rendering scene scales up, the performance bottleneck gradually shifts from splatting to LoD search.
Our experiments show that LoD search can take up to 70\% of the overall execution time in \Sect{sec:bg:perf}.
Thus, it is important to address both stages to achieve high performance across various scales.

\paragraph{LoD Search.} We first dissect the bottlenecks in the LoD search. The bottlenecks of the LoD search are mainly from two aspects: imbalanced workload and irregular memory access.
Algorithmically, PBNR uses a LoD tree to represent scene attributes such as geometry and color. 
The LoD tree itself is imbalanced because each tree node has an unfixed number of child nodes.
Such irregularity makes its tree traversal hard to parallelize on today's GPUs, often resulting in imbalanced workloads across threads.
In addition, the irregular tree traversal often introduces irregular memory accesses, leading to pipeline stalls and costly DRAM accesses~\cite{feng2022crescent, xu2019tigris}.

To address these two issues, we propose \tree, a novel data structure that can translate a canonical LoD tree into a subtree-based data structure while preserving the hierarchical relationship in the original LoD tree (\Sect{sec:algo}).
Unlike other tree structures, e.g., kd-tree~\cite{bentley1975multidimensional} or octree~\cite{meagher1982geometric}, \tree ensures balanced workloads across threads by restricting subtrees to similar sizes, allowing one thread to process one subtree at a time.
In addition, our \tree data structure inherently groups closely visited tree nodes together, preserving high spatial locality. 
With our co-designed traversal algorithm, the tree traversal naturally poses ``structures'' on memory accesses and turns irregular DRAM accesses into streaming ones.
Note that, \tree does not change the semantics of the algorithm and generates bit-accurate results as the canonical LoD tree.

While \tree traversal addresses static workload imbalance, as the camera pose moves during rendering, the workload could vary dynamically throughout execution.
Conventional tree-traversal accelerators~\cite{feng2022crescent, pinkham2020quicknn, xu2019tigris, chen2023parallelnn} fail to address dynamic workload changes because they often adopt offline scheduling, assigning every thread with an equal workload subtree.
To address the dynamic workload, we propose our architectural support, \arch, which supports dynamic scheduling to adapt to PBNR’s view-dependent workloads with a novel subtree cache for tree node lookup (\Sect{sec:arch:core}).

\paragraph{Splatting.}
As the second major bottleneck in PBNRs, splatting suffers from severe warp divergence. 
Rather than adopting advanced Gaussian-tile intersections~\cite{lee2024gscore} to mitigate this issue, we propose a simple yet principled algorithm-hardware co-design to eliminate warp divergence completely. 
Our key observation is that adjacent pixels typically integrate similar sets of Gaussians. 
Based on this insight, we group pixels into small blocks and approximate the Gaussian transparency check by performing this checking at the group-level rather than the per-pixel level. 
This way, all pixels within a group share the same Gaussian integration list, effectively removing warp divergence from splatting.

\paragraph{Results.} Our experiments show that, \proj achieves 3.9$\times$ speedup and 98\% energy savings against a off-the-shelf mobile GPU with our algorithm-hardware co-design.
Compared against existing accelerators, our design can achieve 1.8$\times$ speedup and 54\% energy savings with a comparable area.

Our contributions are summarized as follows:
\begin{itemize}
    \item We introduce a novel data structure, \tree, along with its co-designed algorithm, that tames the imbalance workloads and irregular DRAM accesses in LoD search with bit-accurate results as the canonical LoD tree.
    \item We propose \proj, the first-of-its-kind accelerator for large-scale PBNR, to address dynamic imbalance in \tree traversal and warp divergence in splatting.
    \item \proj achieves 3.9$\times$ speedup and 98\% energy savings against a mobile GPU. 
    With a similar chip area, \proj achieves 1.8$\times$ speedup and 54\% energy savings against a state-of-the-art PBNR accelerator, GSCore.
\end{itemize}

\section{Background and Motivation}
\label{sec:bg}

\begin{figure*}[t]
    \centering
    \includegraphics[width=0.9\textwidth]{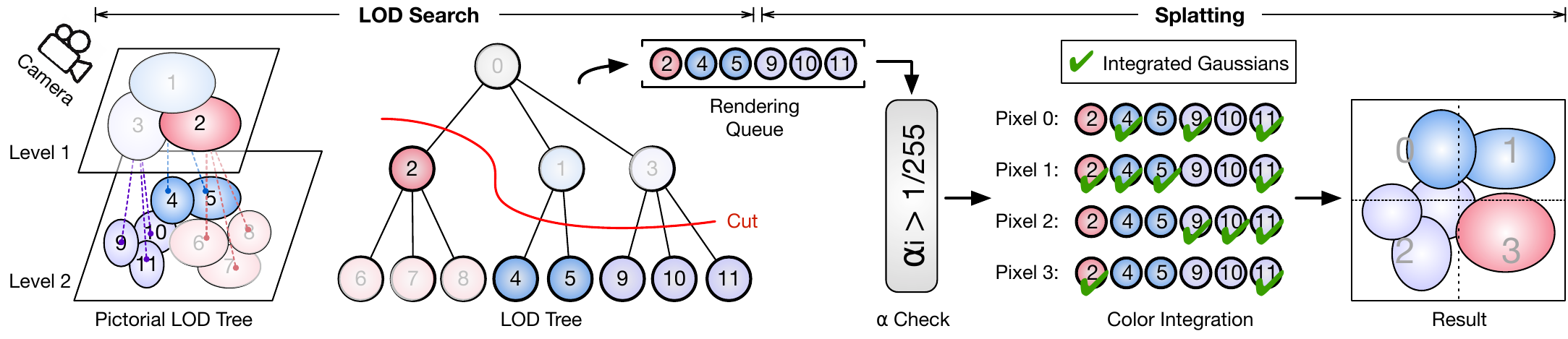}
    \caption{
    An example of the scalable PBNR pipeline primarily consists of two steps: LoD search and splatting. In LoD search, Gaussians at the defined LoD are selected; the selected Gaussians are known as a ``cut'' of the LoD tree. The selected Gaussians first check the intersections with pixels and then ``splats'' on the screen. In the splatting stage, the green marks highlight the integrated Gaussian of each pixel. On GPUs, this sparse color integration leads to warp divergence.
    }
    \label{fig:gs_pipeline}
\end{figure*}

\subsection{Scalable PBNR}
\label{sec:bg:gs}

\paragraph{PBNR.} Recent advancements in deep learning have revived point-based rendering~\cite{wu2024recent, chen2024survey}.
Instead of manually defining the attributes of each rendering primitive, i.e., Gaussians, PBNR leverages the automatic differentiation in deep learning to learn Gaussian attributes~\cite{kerbl20233d, kerbl2024hierarchical}.
Compared to other neural rendering techniques, such as NeRF-based algorithms~\cite{mildenhall2021nerf, gao2022nerf}, PBNR is more efficient by directly rasterizing the rendering primitives, a.k.a., Gaussians, onto the screen, avoiding compute-intensive ray sampling~\cite{pharr2023physically, mildenhall2021nerf}.

However, as the rendering scenes expand, directly rasterizing all Gaussians quickly becomes compute-intensive, since the rendering workload is proportional to the number of rendered Gaussians.
To render scenes at any scales, prior works introduce a hierarchical representation to manage Gaussians~\cite{kerbl2024hierarchical, ren2024octree}.
Overall, the PBNR algorithm consists of two main steps: \textit{LoD search} and \textit{splatting}.

\paragraph{LoD Tree.} Having a hierarchical representation has two main purposes.
First, a hierarchical representation allows for the quick identification of the Gaussians inside the field of view, eliminating redundant computation.
Second, it supports rendering at an appropriate level-of-detail (LoD).
Extremely fine-grained LoD is often an overkill.
For instance, when splatted Gaussians are smaller than the dimension of a single pixel, the finer details are lost since each pixel has one color.

In PBNR, this hierarchical representation is often implemented as a tree structure called \textit{LoD tree}, where each tree level represents a certain detail granularity.
Every tree node~\footnote{We use ``Gaussian'', ``node'', and ``tree node'' interchangeably since there is a one-to-one mapping between them.} is a single Gaussian with an unfixed number of child nodes.
The Gaussians in lower levels are generally smaller and provide finer granularity, as shown in \Fig{fig:gs_pipeline}.
The child nodes represent increasingly detailed textures of its parent node.
For instance, node $2$ has three child nodes: $6$, $7$, and $8$.
These three nodes together represent the finer detail of the node $2$.
In the actual LoD tree from the HierarchicalGS dataset~\cite{kerbl2024hierarchical}, the tree height reaches 24 levels, with one parent node over $10^3$ child nodes.

\paragraph{LoD Search.} For a specified LoD, the appropriate tree level for rendering is determined individually for each Gaussian. 
We call this step \textit{LoD search}.
During rendering, the tree is traversed from top to bottom. 
At each node, we assess whether the projected dimension of the Gaussian at that node is larger than the defined LoD, while the projected dimensions of all its child nodes are smaller. 
If this condition is met, the node and all its child nodes are selected for rendering.
The final rendered Gaussian is an interpolation between them to ensure a smooth fit to the target LoD.
In \Fig{fig:gs_pipeline}, with the camera posed near the left side of the scene, nodes $1$ and $3$ appear too coarse-grained, thus, the subtree nodes of nodes $1$ and $3$ are selected for rendering. 
In contrast, node $2$, which is far from the camera, has a projected dimension that is already smaller than the defined LoD, so node 2 itself is selected.
Eventually, the selected Gaussians form a ``cut'' that separates the top and bottom of the LoD tree, as shown in \Fig{fig:gs_pipeline}.

\paragraph{Splatting.} 
Once the ``cut'' is determined, all the selected Gaussians form a rendering queue.
The second step is to splat the selected Gaussians onto a screen.
Splatting first identifies which Gaussians intersect with each pixel.
Next, the intersected Gaussians are sorted by depth, from the nearest to the farthest. 
Lastly, each pixel integrates the colors of the intersected Gaussians in sorted order to produce its final value.

Notice that, since each pixel typically intersects only a subset of Gaussians from the rendering queue, different pixels would integrate different Gaussians.
The green marks in \Fig{fig:gs_pipeline} highlight the integrated Gaussians of each pixel.
On GPUs, each thread is responsible for one pixel, and threads in a warp execute in lockstep.
In the color integration, GPUs would mask those threads that do not require color integration for particular pixels. 
Due to this divergent color integration process, splatting often introduces warp divergence on existing GPUs.

\subsection{Performance Bottlenecks}
\label{sec:bg:perf}

\begin{figure}[t]
\centering
\begin{minipage}[t]{0.48\columnwidth}
  \centering
  \includegraphics[width=\columnwidth]{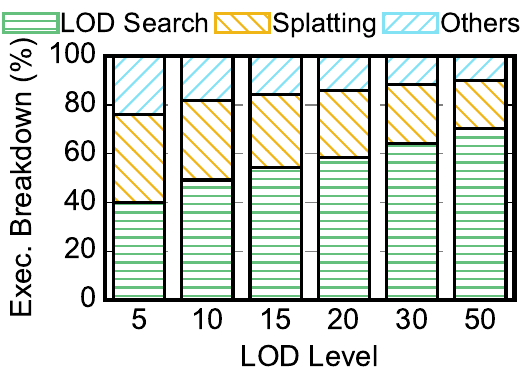}
  \caption{Normalized execution breakdown of PBNR across different LoDs.}
  \label{fig:exec_time}
\end{minipage}
\hspace{2pt}
\begin{minipage}[t]{0.48\columnwidth}
  \centering
  \includegraphics[width=\columnwidth]{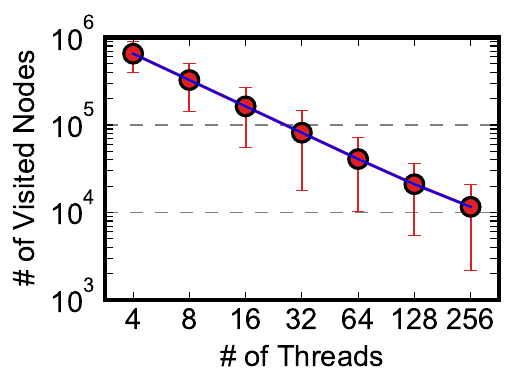}
  \caption{The workload variation as the number of GPU threads increases.}
  \label{fig:imbalance}
\end{minipage}
\end{figure}

Here, we list three performance bottlenecks in PBNRs.

\paragraph{Bottleneck 1.} 
\textit{Two stages, LoD search and splatting, dominate the overall execution time in scalable PBNR.} 
\Fig{fig:exec_time} shows the execution breakdown of PBNR across various rendering scenarios on a Nvidia mobile Ampere GPU~\cite{orinsoc}. 
When the LoD level is low, the percentage of the execution time is roughly the same between LoD search and splatting.
As the camera pose moves farther from the scene and captures a wider view of it, the LoD search phase also becomes the primary contributor to execution time (up to 70\%), compared to splatting.
Nevertheless, LoD search and splatting contribute to 85\% of the overall execution time, on average.

However, existing PBNR accelerators~\cite{lee2024gscore, feng2024potamoi, lee2025vr, li2025uni, ye2025gaussian, he2025gsarch} primarily focus on accelerating splatting, rather than LoD search. 
This paper pinpoints the shift in the primary bottleneck as scene complexity scales. 
Here, we propose an algorithm-architecture co-design to address the LoD search and splatting together.

\paragraph{Bottleneck 2.} 
\textit{LoD search suffers from workload imbalance at runtime due to the dynamic irregularity of the LoD tree.}
Although conventional tree-like structures, such as kd-tree~\cite{bentley1975multidimensional} or octree~\cite{meagher1982geometric}, are statically balanced by design, they still suffer from dynamic workload imbalance due to the irregularity of tree traversal.
For LoD trees, the situation is even worse: the number of child nodes varies depending on scenes. 
Straightforward subtree partitioning, i.e., one thread per subtree, would lead to severe workload imbalance.

\Fig{fig:imbalance} illustrates the workload imbalance across threads when each thread is assigned to a subtree. 
The workload is quantified by the number of visited nodes.
Results show that, with 64 threads, the standard deviation of workload is $3.1\times10^4$ with an average workload of $4.1\times10^4$.
A high workload imbalance could lead to low GPU utilization due to warp divergence and irregular memory accesses.
To avoid this, the existing solutions are to simply apply exhaustive searches to all tree nodes~\cite{kerbl2024hierarchical, ren2024octree}.

\paragraph{Bottleneck 3.} 
\textit{Existing splatting dataflow introduces severe warp divergence due to the sparsity of Gaussian color integration.}
\Sect{sec:bg:gs} describes the color integration process in splatting (see \Fig{fig:gs_pipeline}).
Due to the ``lockstep'' execution paradigm on GPUs, threads that do not intersect a given Gaussian are masked.
For example, in \Fig{fig:gs_pipeline}, if a warp consists of four threads, with one thread for one pixel.
On average, only half of the pixels integrate a particular Gaussian, and the GPU utilization drops to 50\% due to warp divergence.
During the actual splatting stage, our experiments show that the GPU utilization could be as low as 31\%.
Thus, improving the utilization of compute units is critical for rendering efficiency.

\paragraph{Summary.} To address the above performance challenges, we propose \proj to accelerate the LoD search and splatting in PBNR holistically. 
Specifically, we propose \tree traversal (\Sect{sec:algo}) to eliminate the workload imbalance in the LoD search with architecture support (\Sect{sec:arch:core}). 
Meanwhile, we propose a clean-slate accelerator design to address warp divergence in splatting (\Sect{sec:arch:spcore}).


\section{\tree Traversal}
\label{sec:algo}

This section first describes our tree traversal algorithm that streamingly processes LoD trees in parallel (\Sect{sec:algo:overview}).
We then describe our method that converts a canonical LoD tree into our proposed data structure \textit{without} altering algorithmic semantics (\Sect{sec:algo:subtree}).

\subsection{Algorithm}
\label{sec:algo:overview}

\paragraph{Objective.} 
To address the performance bottlenecks in \Sect{sec:bg:perf}, the ideal LoD tree traversal must meet the following design requirements: first, the algorithm should be parallelizable, distributing workloads evenly across threads; 
second, it should be fully streaming, ensuring that any data brought from off-chip is loaded contiguously on-chip with no intermediate data write-back.

\begin{figure}[t]
    \centering
    \includegraphics[width=\columnwidth]{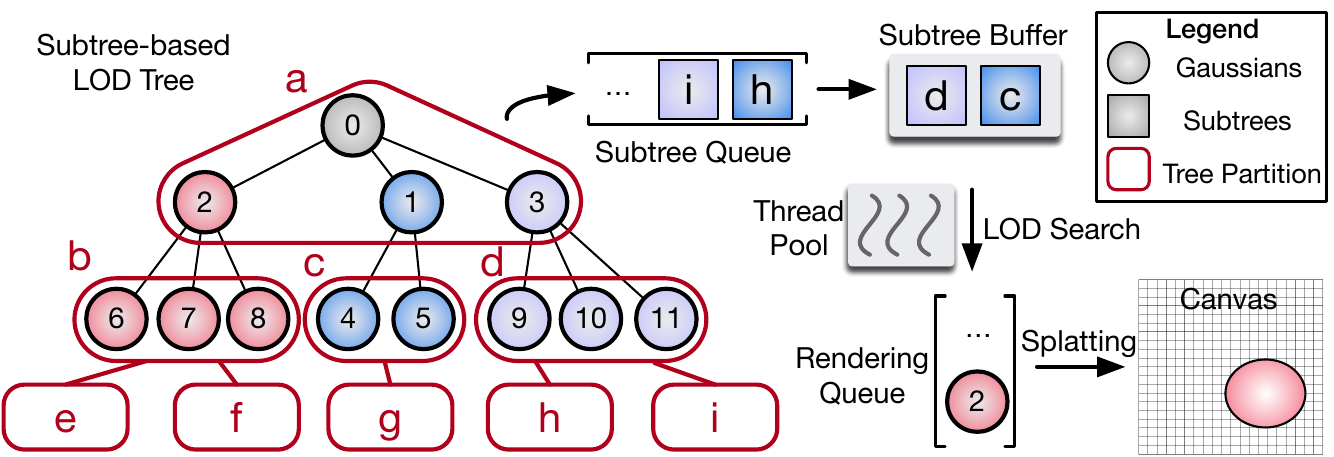}
    \caption{
    Our designed rendering pipeline of PBNR. We first split the LoD tree into small subtrees which still preserve the dependencies of the original LoD tree.
    Tree traversal is then performed at a subtree granularity, with each available thread in the thread pool responsible for one subtree.
    The algorithm terminates when a ``cut'' is obtained across the tree.
    }
    \label{fig:algo}
\end{figure}

\paragraph{Idea.} 
The overall algorithm procedure is illustrated in \Fig{fig:algo}. 
We first partition the entire LoD tree into small comparable-size subtrees while preserving the hierarchical relationships in the original LoD tree. 
For example, nodes $6$ and $7$ remain child nodes of node $2$, and subtree $a$ remains the parent of subtree $b$.
We do not restrict subtree shape, meaning one subtree can include Gaussians within one level or across multiple levels.
However, we set each subtree size to be less than a size limit, $\tau_s$, to ensure that every subtree has a similar workload.
We describe subtree partitioning in \Sect{sec:algo:subtree}.
\tree partitioning is done completely offline with no runtime overhead.

Once the subtree-based LoD tree (\tree) is constructed, we perform a breadth-first search (BFS) to traverse the \tree to find the selected Gaussians on the ``cut'' for subsequent splatting. 
As shown in \Fig{fig:algo}, we start with the top subtree $a$, where node $2$ meets the LoD requirement and is added to the rendering queue. 
However, nodes $1$ and $3$ do not meet the LoD requirement, the algorithm further traverses subtrees $c$ and $d$, which are child nodes of nodes $1$ and $3$.
Our algorithm then loads subtrees $c$ and $d$ into the subtree buffer for LoD search. 
If the Gaussians in subtrees $c$ and $d$ still do not meet the LoD requirement, a deeper tree traversal is necessary, potentially requiring subtrees $i$ and $h$ to be added to the subtree queue.
The tree traversal stops when there is a clean ``cut'' (shown in \Fig{fig:gs_pipeline}) across the \tree.

\paragraph{Streamingly Processing.} 
In \Fig{fig:algo}, our algorithm sequentially adds subtrees that require further traversal to the subtree queue on demand during LoD search.
All tree nodes within a subtree are stored continuously in DRAM, thus, the off-chip memory access becomes streaming.
Available threads in the thread pool then retrieve subtrees from the queue to perform subtree searches in parallel.
Because the subtrees are roughly the same size,  the workloads across threads are balanced.
Once \tree traversal is done, the splatting step then renders all selected Gaussians.


\subsection{\tree Partitioning}
\label{sec:algo:subtree}

Next, we describe \tree partitioning, which consists of two main steps: \textit{initial partitioning} and \textit{subtree merging}.

\begin{figure}[t]
    \centering
    \includegraphics[width=\columnwidth]{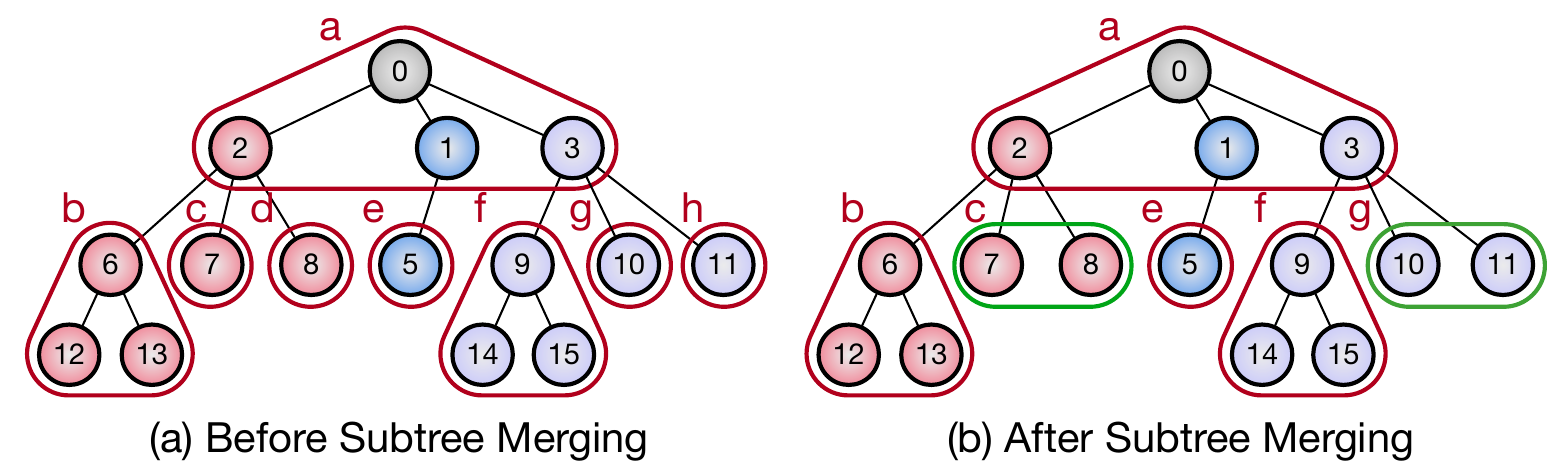}
    \caption{Comparison \tree before and after subtree merging. Before subtree merging, the subtree sizes still vary, leading to workload imbalance.}
    \label{fig:merge_tree}
\end{figure}

\begin{figure*}[t]
\centering
\includegraphics[width=0.9\textwidth]{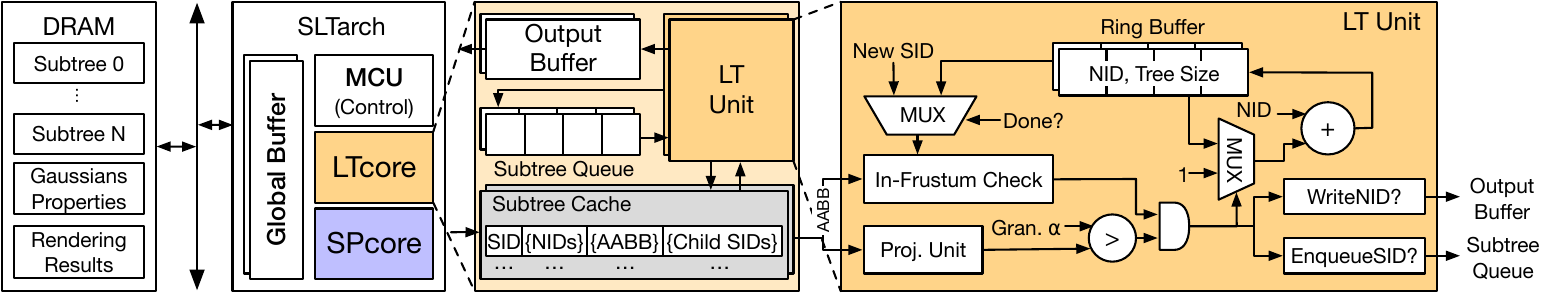}
\caption{The overall \proj architecture design. Our design integrates a LoD search accelerator (\arch) and a splatting accelerator (\spcore). \arch executes LoD search while \spcore supports splatting. Subtrees are initially stored off-chip in BFS order and accessed on a subtree basis. The on-chip global buffer is double-buffered and reads the input data for \spcore. The yellow and purple blocks highlight our key architectural contributions, \arch and \spcore, respectively.}
\label{fig:arch}
\end{figure*}

\paragraph{Initial Partitioning.}
Our partitioning algorithm begins with a BFS traversal from the top of the LoD tree and group tree nodes, as described in \Alg{algo:subtree}. 
When the cumulative number of traversed nodes exceeds the defined subtree size limit, $\tau_s$, we group the traversed nodes into a subtree, $s_j$, and find their immediate child nodes, $N_{child}$. 
These immediate child nodes, $N_{child}$, then become the new roots of their respective LoD subtrees (enqueued in $Q$ in \Alg{algo:subtree}), and BFS is performed individually on each of these new tree roots in $Q$. 
This subtree partitioning process is repeated until all nodes in the original LoD tree are classified into subtrees (\Fig{fig:merge_tree}a).

\begin{algorithm}[t]
\caption{Algorithm of \tree Partitioning}\label{algo:subtree}
\KwData{a list of tree nodes $ N $, tree size limit $\tau_s$ }
\KwResult{a list of subtree $ S $}
$Q \leftarrow N.\text{dequeue()}$, $S_{init}\leftarrow \{\ \}$\;
\While{$Q$ is not empty}{
  $i \gets Q.\text{dequeue()}$\;
  $s_j, N_{child} \gets \text{BFS(}i, N, \tau_s\text{)}$\;
   $S_{init}\text{.push(}s_j\text{)}$\;
  \For{ $n_i$ in $N_{child}$ }{
    $ Q\text{.enqueue(} n_i \text{)} $\;
  }
}

$s_{merge} \leftarrow \{\ \}$, $S\leftarrow \{\ \}$\;

\For{$s \in S_{init}$}{
    \If{ $s.$\text{parent()} is $s_{merge}$.\text{parent()} \textbf{and} $s.\text{size()} \le \tau_s/2$ \textbf{and}
     $s.\text{size()} + s_{merge}.\text{size()} \le \tau_s$}{
        $s_{merge}\leftarrow \text{merge(}s_{merge}, s\text{)}$\;
    }
    \Else{
        $S\text{.push(}s_{merge}\text{)}$\;
        $s_{cur} \leftarrow s$\;
    }
}
\end{algorithm}

Although the initial partitioning divides the original LoD tree into individual subtrees, some subtrees may end up being too small, as shown in \Fig{fig:merge_tree}a. 
For example, with a subtree size limit of 4, subtrees $c$ and $d$ contain only a single node each, which still leads to a workload imbalance between subtrees.

\paragraph{Subtree Merging.} 
To reduce size variation among subtrees, we propose a subtree merging technique. 
Our observation is that some subtrees, e.g., $c$ and $d$ in \Fig{fig:merge_tree}a, can be combined without violating hierarchical relationships.
Therefore, we iterate through the initial partitioned subtrees to identify small subtrees (those subtree sizes are smaller than half of the size limit, $\tau_s/2$) and check if they can be merged with adjacent ones that share the same parent subtree. 
For example, subtree $c$ would check for other small subtrees under the same parent node (node 2) and merge with subtree d.
This merging process is performed in a greedy manner and stopped when the size of the current merged subtree, $s_{merge}$, will exceed $\tau_s$. 
The final \tree, shown in \Fig{fig:merge_tree}b, reduces workload imbalance compared to the initial \tree in \Fig{fig:merge_tree}a.
We quantitatively evaluate the benefit of the subtree merging in \Sect{sec:eval:abl}.





\section{Architectural Support}
\label{sec:arch}

\Sect{sec:algo} explains our \tree traversal that enables static balanced workload across threads and streaming process in LoD search. 
This section introduces our architectural supports that address the dynamic workload imbalance in LoD search and warp divergence in splatting. 
We begin with an overview of our architectural design (\Sect{sec:arch:overview}), and we then explain the key components that support LoD search (\Sect{sec:arch:core}) and splatting (\Sect{sec:arch:spcore}), respectively.

\subsection{Overview}
\label{sec:arch:overview}

\Fig{fig:arch} shows our overall \proj architecture, which consists of a tree traversal core (\arch), a splatting core (\spcore), and a global buffer to store intermediate data.
In our architecture design, \arch is dedicated to the \textit{LoD search} step, while \spcore executes the \textit{splatting} step.

\begin{figure}[t]
    \centering
    \includegraphics[width=0.9\columnwidth]{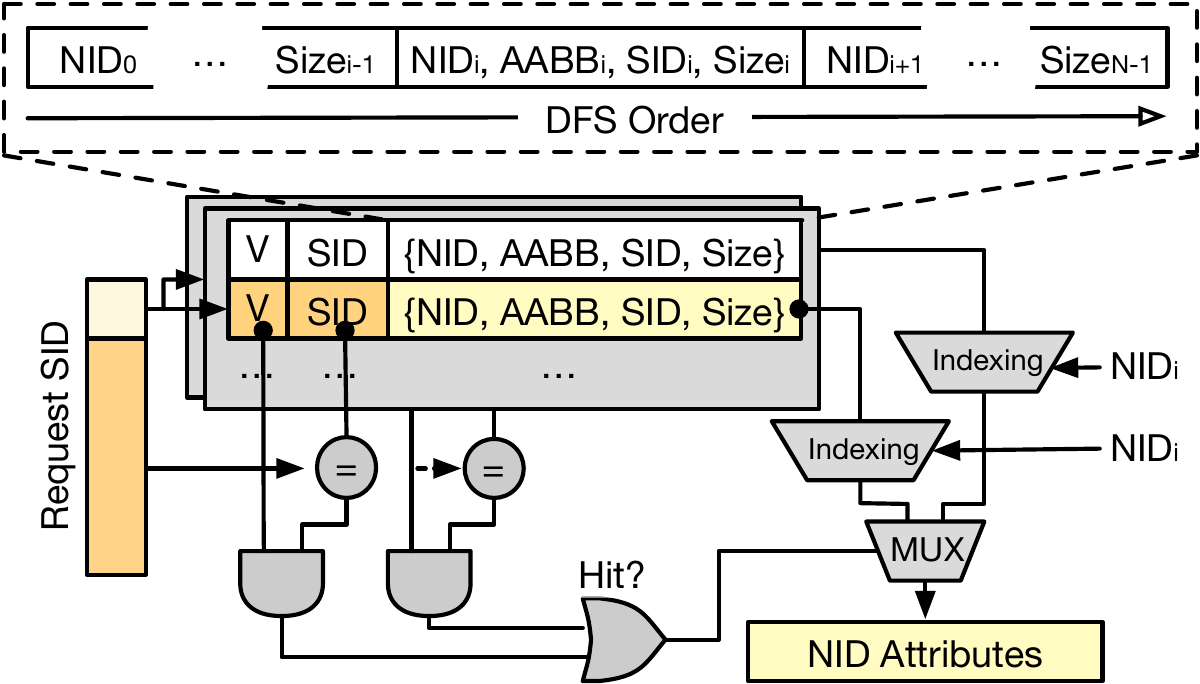}
    \caption{The subtree cache design. Each cache tag is a SID, and each cache entry stores all node attributes needed for a single subtree traversal. Each node attribute can be retrieved by indexing with the NID.
    }
    \label{fig:cache}
\end{figure}

\paragraph{Overall Dataflow.} Our \arch consists of an array of LT units, a subtree queue, a subtree cache, and an output buffer. 
The subtree queue stores the subtree IDs (SIDs) that need to be traversed. 
Each LT unit is responsible for processing one subtree. 
Whenever one LT unit becomes available, it signals the subtree queue to dequeue a new SID. 
The LT unit then traverses the subtree associated with this SID.
Each tree node is accessed from the subtree cache using the node ID (NID), as shown in \Fig{fig:cache}. 
For each tree node, the LT unit checks if it meets the LoD requirement, a.k.a, the cut in \Fig{fig:gs_pipeline}; if satisfied, the NID is written to the output buffer.
The output buffer is double-buffered, one write-back buffer and one filling buffer.
Once the filling buffer is full, the two buffers are swapped, allowing continuous \tree traversal without pipeline stalls.

\begin{figure*}[t]
\centering
\includegraphics[width=0.9\textwidth]{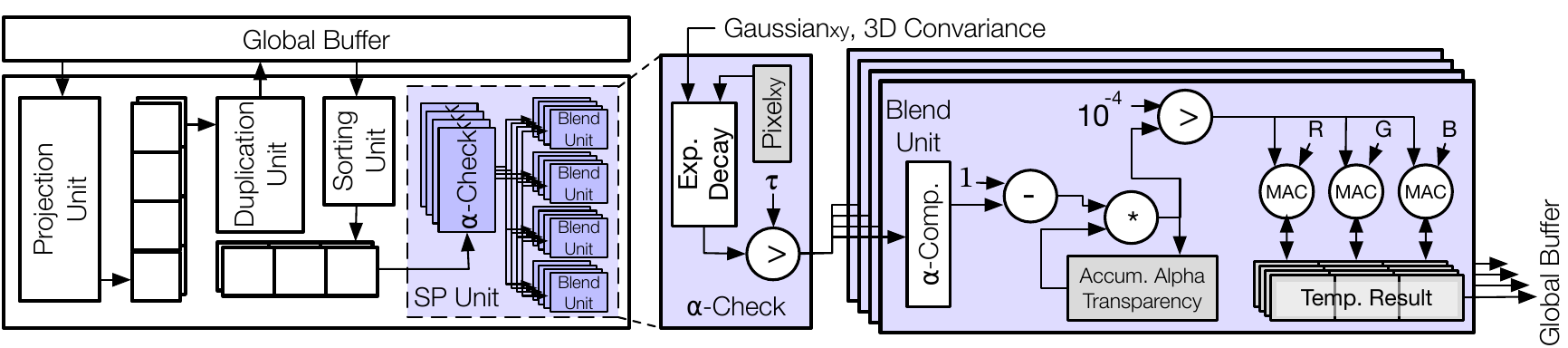}
\caption{The \spcore architecture design. Overall, \spcore consists of a projection unit, a duplication unit, a sorting unit, and a set of SP units. By and large, our design is built upon the hardware design of GSCore~\cite{lee2024gscore}. Our main contribution is the new splatting unit, SP unit, to address warp divergence in splatting. Purple blocks highlight our key architectural contributions.}
\label{fig:spcore}
\end{figure*}

During the splatting step, the global buffer first reads Gaussians, which are required from DRAM.
This global buffer is also double-buffered to hide the latency of data loading.
\spcore then reads data from the global buffer and renders the final image.
Detailed design of \spcore is shown in \Fig{fig:spcore}.

\paragraph{Fully-Streaming \tree Traversal.} 
It is worth noting that, in \tree traversal, all tree nodes within a subtree are stored continuously in DRAM. 
During the execution, we load one subtree entirely into the subtree cache \textit{on demand} at a time.
Thus, our design guarantees that DRAM accesses of one subtree are streaming. 
Here, we do not align the DRAM row boundaries with the subtree size. 
Nevertheless, aligning the DRAM row boundary can boost potentially performance by carefully designing the data layout. 

\subsection{\arch}
\label{sec:arch:core}

\paragraph{Motivation.}
We begin by motivating the need for a new tree traversal core. 
Many prior works have proposed accelerators for tree-based algorithms.
However, they primarily focus on tree structures, such as kd-trees~\cite{feng2022crescent, xu2019tigris, pinkham2020quicknn, feng2025streamgrid}, and octrees~\cite{chen2023parallelnn}.
These existing accelerators rely on offline workload scheduling, which cannot address the dynamic workload imbalance in the LoD tree that arises as the camera moves at runtime.
Moreover, those designs require a dedicated stack per compute unit for tree tracebacks.
In the context of LoD trees, where the number of child nodes varies across the structure, this leads to insufficient or wasted stack buffers in prior accelerators. 
Thus, we propose a co-designed accelerator tightly coupled with our proposed \tree traversal.

\paragraph{LT Unit.} 
\Fig{fig:arch} shows that our \arch has a $2 \times 2$ array of LT units to traverse a \tree proposed in \Sect{sec:algo}.
LT unit is designed to pipeline between different subtree traversals.
Within each LT unit, a small SRAM acts as a ring buffer to store the states of different subtree traversals.
Every cycle, the LT unit checks if the current subtree traversal is complete by comparing the current NID with the current subtree range. If the current NID surpasses the current subtree range, the LT unit requests a new SID from the subtree queue and begins traversing from the top of the new subtree.

To avoid the pipeline stalls of LT units, the subtree queue is separated into two segments. 
One segment stores the SIDs that are already loaded into the subtree cache, while the other contains the unloaded SIDs.
Once the data for an unloaded SID are loaded into the cache, this SID is moved to the loaded segment.
LT units can only access SIDs from the loaded segment.
This design guarantees that all subtrees traversed by LT units are always in the subtree cache, so that the LT units will never be stalled due to cache misses.

During a tree node traversal, the LT unit requests the axis-aligned bounding box (AABB) of the current NID from the subtree cache and checks two conditions.
The first is whether the node lies within the current rendering view frustum, and the second is whether the NID meets the LoD requirement. 
If both conditions are met, this NID is written to the output buffer and skips its remaining subtree, i.e., any nodes beneath this NID.
This skipping is achieved by incrementing the current NID with the remaining subtree size.
If the conditions are not met, the current NID is updated by 1 and continues the remaining subtree traversal.
When the current NID is the leaf node of this subtree, it enqueues its child SID to the subtree queue for further tree traversal.

\paragraph{Subtree Cache.}
Our subtree cache is designed as a 4-way set-associative cache as shown in \Fig{fig:cache}.
Here, we draw a 2-way set-associative cache for illustration purposes.
Each cache entry stores one SID as a cache tag along with all the NIDs associated with this SID. 
In addition to NIDs, a cache entry also includes all NIDs' AABBs, remaining subtree sizes, and their corresponding child SIDs. 
Given that each subtree has a defined size limit, the number of NIDs stored per entry is set to this limit, with zeros padded if the subtree contains fewer nodes than the limit.
All NIDs are stored in a depth-first search order, allowing us to skip unnecessary computations by bypassing the current node’s subtree if the current NID meets the LoD requirement or has no intersection with the frustum.

When replacing a cache entry, we first use the SID to index the cache and check if any subtrees in the entry are complete. 
If a subtree is finished, we directly replace that cache entry with the new one. 
If no subtrees are finished, we stall the cache update.
Given our streaming tree traversal algorithm, once a cache entry is evicted, it will not be reloaded during the rest of the SLTree traversal.
As replacement policies have no impact on performance, we use a round-robin replacement policy.

\subsection{\spcore}
\label{sec:arch:spcore}

\paragraph{Motivation.} 
Recall, in \Fig{fig:gs_pipeline}, due to the per-pixel $\alpha$ check, different pixels would integrate different subsets of Gaussians in the rendering queue.
This sparse color integration introduces warp divergences in the splatting step.
Although quite a handful of accelerators have recently been proposed for PBNR~\cite{lee2024gscore, feng2024potamoi, lee2025vr, li2025uni, ye2025gaussian, he2025gsarch}, few directly address the key bottleneck in splatting: warp divergence (\Sect{sec:bg:perf}). 
For instance, GSCore~\cite{lee2024gscore} introduces a finer-grained Gaussian-tile intersection strategy to reduce false-positive intersections. 
However, this approach introduces non-trivial computational overhead and complicates the overall hardware design. 
Rather, we propose a simple yet effective algorithm-hardware co-design that eliminates warp divergence completely.

\paragraph{Overview.}
\Fig{fig:spcore} illustrates our \spcore architecture design.
Overall, \spcore consists of a projection unit, a duplication unit, a sorting unit, and four SP units.
Note that, our design is built upon the hardware design of GSCore~\cite{lee2024gscore}.
Our main contribution is the new splatting unit, SP unit, to address warp divergence in splatting, which is the second main bottleneck in PBNRs.
We keep the other three components untouched since they are responsible for the computations categorized as ``others'' in \Fig{fig:exec_time} and contribute merely 15\% of the total execution time.
We also simplify the design of the projection unit by using the basic 3-$\sigma$ Gaussian-tile intersection test, instead of precise intersection tests, e.g., Axis-Aligned Bounding Box~\cite{klosowski1998efficient} or Oriented Bounding Box~\cite{gottschalk1996obbtree} tests, which would otherwise increase the hardware complexity. 
Because our SP unit naturally performs finer-grained Gaussian-tile filtering.
Nevertheless, we claim no contribution for these three components.

\paragraph{SP Unit.}
The warp divergence in splatting arises from the fact that different pixels within a warp would integrate different sets of Gaussians.
To eliminate this divergence, our observation is that the transparencies ($\alpha$ value) of a given Gaussian are similar across adjacent pixels.
Leveraging this insight, we split all pixels into $2\times2$ pixel groups.
Instead of evaluating each pixel individually, we compute the transparency of the Gaussian using the center of the pixel group.
If this value falls below the threshold ($\frac{1}{255}$ in \Fig{fig:gs_pipeline}), we can skip the color integration of this Gaussian for the entire pixel group.
In this way, there is no divergence within a pixel group.

Our SP unit design in \Fig{fig:spcore} exploits this insight.
Each SP unit consists of one $\alpha$-check unit and four blending units.
The $\alpha$-check unit computes the transparency of a Gaussian.
If the transparency is low, we stop sending this Gaussian to the four blending units for the remaining color integration.
Note that, computing transparency requires exponent computation, which is compute-heavy.
Here, we avoid such a computation in the $\alpha$-check unit by checking the power of the exponent instead.
\Sect{sec:eval:perf} shows that this simple yet principled hardware augmentation leads to a performance gain compared to GSCore.

\section{Evaluation}
\label{sec:eval}

\subsection{Experimental Setup}
\label{sec:eval:exp}

\paragraph{Hardware Setup.} 
As shown in \Fig{fig:arch}, \proj architecture has two parts: \arch and \spcore.
Our \arch consists of $2 \times 2$ LT units clocked at 1 GHz, a subtree queue with a size of $1\times48$~B, and a double-buffered output buffer with a size of 8~KB. 
Subtree cache is a 4-way associative cache, comprising $4 \times 128$ entries with a total size of 128~KB.
Our \spcore, which is also clocked at 1 GHz, consists of 4 projection units, 4 sorting units and a $2\times 2$ SP units with a 256~KB double-buffered global buffer.
The number of projection units and sorting units in \spcore is the same as the original paper~\cite{lee2024gscore}.
The entire \proj architecture design is developed using an EDA process and synthesized with Synopsys and Cadence tools on TSMC’s 16~nm FinFET technology.
The GPU performance and power are directly measured from a mobile Ampere GPU via the Nvidia power monitor API.
GPU results are scaled to 16~nm using DeepScaleTool~\cite{sarangi2021deepscaletool} to be compatible with our simulation.

SRAM components are generated using the Arm Artisan memory compiler, with power estimated via Synopsys PrimeTimePX with annotated fixed-value switching activities.
The DRAM model in our simulations is based on Micron’s 32 Gb LPDDR4 with 4 channels according to its datasheet~\cite{micronlpddr4}, with energy consumption data sourced from Micron System Power Calculators~\cite{microdrampower}.
The overall energy of random DRAM access and random SRAM access is about 25:1, and non-streaming and streaming DRAM access is about 3:1.
Both numbers are aligned with prior works~\cite{gao2017tetris, Yazdanbakhsh2018GAN}.

\paragraph{Area Overhead.}
\proj introduces negligible area overhead compared to a typical mobile SoC ($>$100 mm$^2$)~\cite{orinsoc, xaviersoc, applea15}, with a total area of 1.90~mm$^2$. \arch and \spcore contribute to 0.14~mm$^2$ and 1.76~mm$^2$.
LT Unit array and subtree cache account for 0.03~mm$^2$ and 0.10~mm$^2$ of the \arch area, respectively.
We also scale GSCore's area down to 16~nm using DeepScaleTool~\cite{sarangi2021deepscaletool} and show that \proj has a similar area against GSCore (1.78~mm$^2$).


\paragraph{Software Setup.} 
We evaluate our technique on a widely adopted PBNR algorithm, \mode{HierarchicalGS}, using the large-scale scene reconstruction dataset: HierarchicalGS~\cite{kerbl2024hierarchical}. 
This dataset includes two scenes, each with six rendering scenarios. 
We do not evaluate the datasets used in GSCore, such as Mip360~\cite{barron2021mip}, Tanks\&Temples~\cite{knapitsch2017}, and DeepBlending~\cite{hedman2018deep}, as they are considered small-scale and less representative of real-world large-scene rendering. 
Unless otherwise specified, we set the subtree size to 32 in the paper.

\paragraph{Baselines.} We compare four baselines in our evaluation.
\begin{itemize}
    \item \mode{GPU}: a mobile Ampere GPU on Nvidia Orin SoC~\cite{orinsoc}.
    \item \mode{GPU+LT}: a mobile SoC integrates an Ampere GPU for splatting and \arch for LoD search.
    \item \mode{GPU+GS}: a mobile SoC integrates an Ampere GPU for LoD search and a GSCore for splatting.
    \item \mode{LT+GS}: this variant replaces \spcore with GSCore. \arch runs LoD search and GSCore runs splatting.
    \item \mode{\proj}: our full-fledged architecture in \Fig{fig:arch}. \arch runs LoD search and \spcore executes splatting.
\end{itemize}

\subsection{Accuracy}
\label{sec:eval:acc}

\begin{table} 
\caption{The rendering quality evaluation between the original algorithm and \proj across three quality metrics.}
\resizebox{\linewidth}{!}{
\renewcommand*{\arraystretch}{1}
\renewcommand*{\tabcolsep}{5pt}
\begin{tabular}{ c|cc|cc|cc } 
\toprule[0.15em]
\multirow{2}{*}{Dataset} & \multicolumn{2}{c|}{PSNR (dB)$\uparrow$} & \multicolumn{2}{c|}{SSIM$\downarrow$} & \multicolumn{2}{c}{LPIPS$\downarrow$} \\ 
& Org. & \proj & Org. & \proj & Org. & \proj \\ 
\midrule[0.05em]
Small-scale & 21.05 & 21.04 & 0.758 & 0.756 & 0.289 & 0.291 \\
Large-scale & 23.51 & 23.50 & 0.784 & 0.782 & 0.316 & 0.318  \\
\bottomrule[0.15em]
\end{tabular}
}
\label{tab:eval}
\end{table}

\Tbl{tab:eval} evaluates the rendering quality between the canonical PBNR algorithm and our modified one with three widely-used quality metrics: PSNR, SSIM, and LPIPS.
Overall, our \proj achieves a similar rendering quality with a marginal accuracy loss.
For instance, on PSNR, \proj drops the quality by 0.01 on average.
Note that, \tree traversal does not alter the semantics of the LoD search.
The main accuracy drop is from the rasterization approximation introduced by \spcore in \Sect{sec:arch:spcore}.

\subsection{Performance Evaluation}
\label{sec:eval:perf}

\begin{figure}[t]
    \centering
    \includegraphics[width=0.95\columnwidth]{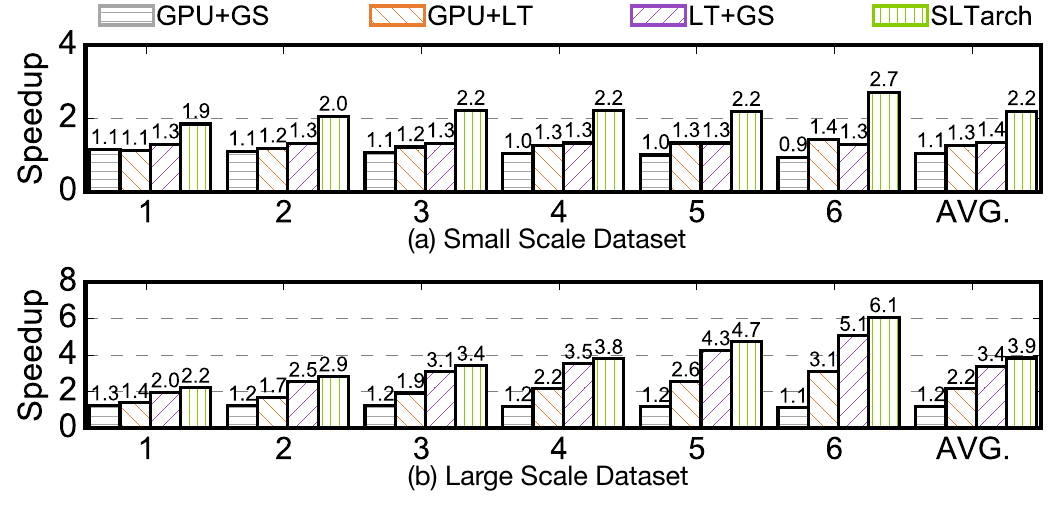}
    \caption{Speedup of different hardware variants over \mode{GPU} baseline on both small-scale and large-scale datasets. Numbers are normalized by \mode{GPU}.
    }
    \label{fig:speedup}
\end{figure}


\paragraph{Performance.} 
\Fig{fig:speedup} shows the speedup of different hardware variants against \mode{GPU} baseline.
For small-scale scenes, \mode{\proj} can merely achieve 2.2$\times$ over 6 scenarios compared to \mode{GPU}. 
However, in large-scale scenes, \mode{\proj} achieves 3.9$\times$ over 6 scenarios against \mode{GPU}, with a maximum speedup up to 6.1$\times$.
In comparison, \mode{GPU+GS} and \mode{GPU+LT} just achieve 1.2$\times$ and 2.2$\times$, respectively.
We also show that \mode{\proj} achieves better performance against \mode{LT+GSCore}.
Our results demonstrate that introducing our \arch into either \mode{GPU+LT} or \mode{\proj} improves overall performance.

\begin{figure}[t]
    \centering
    \includegraphics[width=0.95\columnwidth]{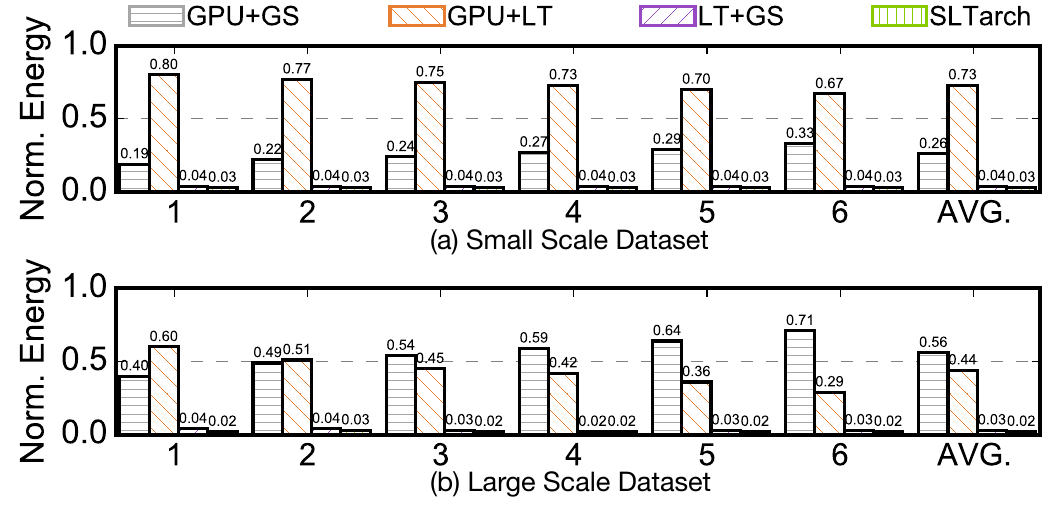}
    \caption{Normalized energy of different variants compared to \mode{GPU} on both small-scale and large-scale datasets.
    }
    \label{fig:energy}
\end{figure}

\paragraph{Energy Savings.} 
\Fig{fig:energy} shows the normalized energy of different hardware variants against the \mode{GPU} baseline. 
All energy values are normalized against the corresponding \mode{GPU} values.
On the small-scale dataset, \mode{GPU+GS} saves 74\% of the total energy compared to \mode{GPU}, while \mode{GPU+LT} only achieves 26\% of the energy savings.
This is because GPU power is the primary energy contributor, and in the small-scale dataset, execution is dominated by splatting rather than LoD search.

For large-scale datasets, \mode{GPU+GS} and \mode{GPU+LT} achieve 44\% and 57\% of the overall energy savings, respectively. 
As shown in \Sect{sec:bg:perf}, the overall execution time is dominated by LoD search.
By integrating GSCore with \arch, \mode{\proj} can save 98\% of the overall energy across both datasets.

\paragraph{DRAM Traffic.} Compared to existing LoD search methods which use exhaustive search to traverse the entire LoD tree to avoid imbalanced workloads across GPU threads.
However, our LoD just search the tree nodes that are within the view frustum and above the ``cut'' in \Fig{fig:gs_pipeline}.
Overall, our LoD search, on average, reduces the DRAM traffic by 76.5\% and 69.6\%, on small-scale and large-scale datasets, respectively.

\subsection{Comparison against Tree Traversal Accelerators}



\begin{figure}[t]
\centering
\begin{minipage}[t]{0.48\columnwidth}
    \centering
    \includegraphics[width=\columnwidth]{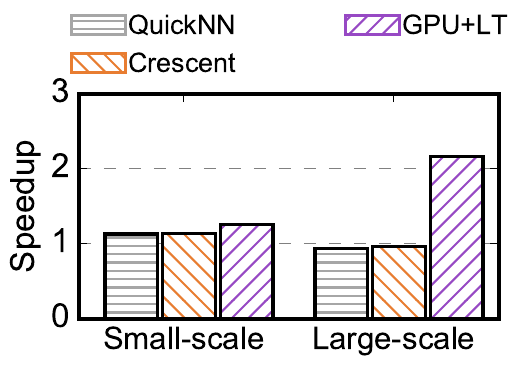}
    \caption{Performance comparison of \mode{GPU+LT} against two prior tree-based accelerators, QuickNN~\cite{pinkham2020quicknn} and Crescent~\cite{feng2022crescent}. 
    Numbers are normalized against \mode{GPU}.
    }
    \label{fig:tree}
\end{minipage}
\hspace{2pt}
\begin{minipage}[t]{0.48\columnwidth}
  \centering
  \includegraphics[width=\columnwidth]{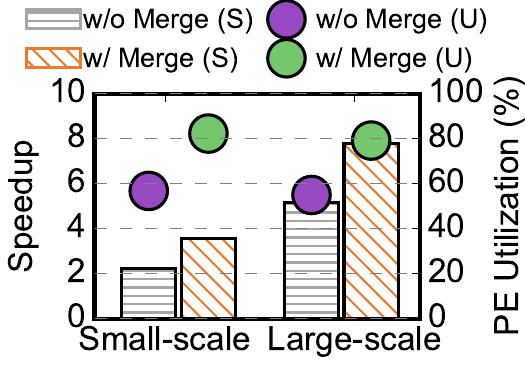}
  \caption{Ablation study of LoD search with and without subtree merging in \Sect{sec:algo:subtree}. 
  `S' and `U' represent speedup and PE utilization.
  Only the performance of the LoD search is shown here.
  }
  \label{fig:ablation}
\end{minipage}
\end{figure}

We also compare the efficiency of our \tree traversal and \arch against two state-of-the-art kd-tree accelerators, QuickNN~\cite{pinkham2020quicknn} and Crescent~\cite{feng2022crescent}.
For a fair comparison, we configure the GPU to execute the splatting stage, while different tree-based accelerators perform the LoD search across all hardware variants, using the same number of PEs.
The performance numbers are normalized against the \mode{GPU} baseline.

Overall, our \mode{GPU+LT} achieves better speedup for two key reasons.
First, kd-tree traversal is inherently ill-suited for LoD search due to memory access irregularity.
Second, both QuickNN and Crescent designs introduce unnecessary computations, such as loading/storing data to the local stack, to accommodate tree tracebacks.
However, these operations are not required for LoD search.

\subsection{Ablation Study}
\label{sec:eval:abl}

\Fig{fig:ablation} presents the ablation study of LoD search with and without subtree merging.
Here, we only show the performance of the LoD search. 
All values are normalized to the \mode{GPU} baseline.
Overall, \mode{arch} without subtree merging achieves 2.3$\times$ and 5.2$\times$ speedup on small-scale and large-scale scenes, respectively.
By applying subtree merging, \mode{arch} further boosts performance to 3.6$\times$ and 7.8$\times$ speedup on small-scale and large-scale scenes, respectively.




\section{Related Work}

\paragraph{PBNR Acceleration.} 
With the growing popularity of PBNRs~\cite{fan2023lightgaussian, fang2024mini, kerbl20233d, kerbl2024hierarchical, huang2025seele}, there is increasing interest in dedicated accelerators for PBNR~\cite{lee2024gscore, feng2024potamoi, lee2025vr, li2025uni, ye2025gaussian, he2025gsarch, feng2025lumina, zhang2025streaminggs, feng2025streamgrid, lin2025metasapiens}.
For instance, VR-pipe~\cite{lee2025vr} augments the existing GPU rasterization pipeline.
GSArch~\cite{he2025gsarch} support PBNR model training.
However, prior work has largely focused on the splatting stage while overlooking the significance of LoD search. 
This study proposed an algorithm-architecture co-designed system to address PBNR scalability.

\paragraph{Tree Traversal Acceleration.} 
Most tree traversal accelerators focus on kd-tree or octree traversal for tasks like k-nearest neighbor search or data compression~\cite{feng2022crescent, chen2023parallelnn, xu2019tigris, pinkham2020quicknn}. 
Our work focuses specifically on LoD tree traversal for rendering, tackling the challenges posed by irregular tree structures. 
Meanwhile, it has the potential to be applied to other irregular tree traversal tasks as well.
\section{Conclusion}

\proj introduces an algorithm-architecture co-design to tackle workload imbalance and memory irregularity in PBNR, taking one step towards scalable PBNR.
The core idea of \proj is to impose ``structure'' on irregular tree traversal and approximate the splatting to reduce warp divergence.
This way, we improve the locality of PBNR and hardware efficiency with minimal hardware support.
\section{Acknowledgements}

This work was supported by the National Natural Science Foundation of China grants (62222210 and 62402312).
This work was also supported by Shanghai Qi Zhi Institute Innovation Program SQZ202316.

\bibliographystyle{ieeetr}
\bibliography{references}

\end{document}